\documentclass[12pt]{article}
\begin{document}
\baselineskip=16pt

\title
{\begin{Huge}
Severe Challenges In Gravity Theories\\
\end{Huge}
\vspace{0.6in} }
\author{Yuan K. Ha\\ Department of Physics, Temple University\\
 Philadelphia, Pennsylvania 19122 U.S.A. \\
 yuanha@temple.edu \\    \vspace{.1in}  }
\date{July 1, 2011}
\maketitle
\vspace{.6in}
\begin{abstract}
\noindent
Gravity is specifically the attractive force between two masses separated at a distance.
Is this force a derived or a fundamental interaction? We believe that all fundamental interactions
are quantum in nature but a derived interaction may be classical. Severe challenges have appeared in
many quantum theories of gravity. None of these theories has thus far attained its goal in quantizing
gravity and some have met remarkable defeat. We are led to ponder whether gravitation is intrinsically
classical and that there would exist a deeper and structurally different underlying theory which would
give rise to classical gravitation, in the sense that statistical mechanics, quantum or classical,
provides the underlying theory of classical thermodynamics.\\
\end{abstract}

\newpage
\noindent
{\bf 1 \hspace{.1in} Introduction}\\

\noindent
In classical gravity, coordinate space is curved and momentum space is flat. This situation would continue
in quantum gravity as far as one follows the standard rules of quantum mechanics and general relativity.
The discrepancy in the nature of coordinate space and momentum space is the ultimate source of conflict
between the two successful theories [1]. In quantum mechanics, the basic cornerstone is wave-particle duality.
Particles have demonstrated wavelike behavior and waves have shown to possess particle properties. Even 
large molecules with significant mass and complexity have demonstrated the basic wave-particle duality [2].
Particles are characterized by energy and momentum; whereas waves are specified by a wavelength - a spatial
extension, and a frequency related to time. This immediately involves two descriptions, one in momentum space 
and one in coordinate space of spacetime. We believe that the two descriptions are equivalent in general.
In addition, there is a Hilbert space in either representations for linear superposition of wave functions.
Linear superposition of wave functions is another cornerstone of quantum mechanics.\\

\vspace{.1in}
\noindent
{\bf 2 \hspace{.1in} A New Physical Reality}\\

\noindent
Gravity is always an attractive force between two masses separated by a distance. It is solely an effect
in spacetime. Two masses in momentum space do not attract each other according to the inverse square law
if they are separated by a distance in momentum space. There is no such effect. This is why coordinate space
is curved and momentum space is flat. With gravity, one can see that coordinate space and momentum space 
are {\it not} equivalent. This is the new physical reality. In quantum mechanics, the connection between
coordinate space and momentum space is achieved with the Fourier transform. Fourier transform is the exact
tool for realizing wave-particle duality. However, such a powerful tool works only in flat space! Flat space
is compulsory for summing up the Fourier components according to the basic properties of sine and cosine
in order to construct wave functions and quantum fields. Fourier transform cannot be defined in curved space
in mathematics, therein lays the real difficultly. What does this mean for wave-particle duality for gravity?
How can one construct a theory of quantum gravity without this very principle?\\

It is therefore impossible to combine general relativity and quantum mechanics at the most fundamental level.
This is in contrast to the Dirac equation in which it is possible to combine special relativity and quantum
mechanics at a certain level since both are based on flat spacetime. The result is Dirac's relativistic
quantum mechanics and its prediction of spin-1/2 antiparticles. This is usually hailed as the reconciliation
of special relativity and quantum mechanics. Even here, we note that the reconciliation is only partial, 
for no one has ever imagined, let alone produced a quantum theory of special relativity in which the Planck
constant $\hbar$ is to appear in the Lorentz transformation. The presence of the Planck constant in a set of
equations is the real hallmark of a quantum theory. Discreteness of eigenvalues, usually accepted as evidence
of quantum behavior, is secondary since classical systems can also produce discrete eigenvalues. This `quantum
theory of special relativity' is supposed to be the zero-gravity limit of a quantum theory of general relativity.
One can easily see that this logical alternative is not possible because the Lorentz transformation is nothing
but a pseudo-rotation in four-dimensional Minkowski spacetime, a purely mathematical transformation. If the limit
of a proposed theory is not consistent, how then can one achieve a full quantum gravity theory with general
coordinate transformation?\\

Thus there is no real reconciliation of spacetime and quantum theory even at the special relativity level.
The reality is that motion and quantum are incompatible from the beginning. Trajectory is not a notion in
quantum mechanics. In any case, a Lorentz transformation would immediately take an event outside the Planck
scale so as to render the quantum gravity domain of $10^{-33}$ cm irrelevant. Translation in spacetime is
therefore not compatible with the meaning of quantum gravity. If a true quantum gravity theory does not deal
with translation, then quantizing general relativity with special relativity as a limit is a contradiction.\\

Matter is constructed out of spin-1/2 fermions and fermions are naturally incorporated in the Dirac equation.
This is fortunate because spinor representations exist in $SO(3,1)$, the group of Lorentz transformations. 
In general relativity, there is no spinor representation in $GL(4,R)$, the group of general coordinate transformations.
Hence fermions do not exist intrinsically in a curved spacetime. At most they can exist in the tangent space
of a curved spacetime because tangent space is flat. Thus fermions must undergo free fall in order to stay as
free particles. They cannot remain at rest in strong gravity without being deformed. To include fermions in a
curved spacetime, the vierbein formalism is needed to achieve a minimum coupling of gravity to fermions in the
Dirac equation by following a gauge invariance principle. This is another difficulty in combining quantum theory
and general relativity when matter is included. The existence of spinor representations is crucial for any
alternative theory of gravity.\\

\vspace{.2in}
\noindent
{\bf 3 \hspace{.1in} The Meaning Of Quantum Gravity}\\

\noindent
Quantum gravity is originally intended to achieve a microscopic theory of gravity similar in vein to
quantum electrodynamics, using the macroscopic variables of spacetime such as the metric tensor $g_{\mu\nu}$
and the connection $\Gamma^{\lambda}_{\mu\nu}$ [3]. This is the covariant quantization approach. Following
the rules of quantum mechanics, one expects to extract the quantum nature of gravity such as superposition,
interference, uncertainty relations and quantum fluctuations of spacetime by treating the metric as a
quantum operator in a suitable Hilbert space. However, this very construction is performed in flat spacetime
because quantum field theory is valid only in a fixed and flat background spacetime. All quantum field theories
of gravity constructed in this approach require a massless spin-2 particle which is the graviton. The graviton
is the particle which corresponds to the weak field perturbation on the flat background spacetime. When Feynman
rules are applied to processes involving the graviton, however, the resulting quantum gravity theory is always
divergent at high energies and becomes nonrenormalizable [4]. No quantum field theory of gravity involving the
graviton has been successful in any prediction of physics in strong gravity.\\

Thus the spin-2 graviton constructed in every quantum field theory of gravity necessarily propagates in a flat
spacetime. This is obviously contradictory to the background independence of Einstein's gravity. The existence
of the graviton itself in nature remains to be seen. It is physically impossible to detect a {\it single} graviton
because of its very low frequency and hence its energy [5]. Gravitons have long wavelengths of many kilometers
and in order to achieve a reasonable probability of capture of a single graviton the detector must be enormous 
in size and comparable to its wavelength. In that case, the detector would become physically too massive and 
collapse into a region less than its Schwarzschild radius and effectively become a black hole. The graviton
would be lost, making the situation quite impossible to verify a quantum field theory of gravity based on the
graviton.\\

On the other hand, a nonperturbative total quantization of gravity without a background spacetime has been developed.
This the the canonical quantization approach. It is based on the Hamiltonian formulation of general relativity
using the metric and curvature as generalized variables. The result is the Wheeler-Dewitt equation [6]. It has 
the reputation of being ill-defined. The Hamiltonian does not involve time derivative but rather a constraint on
the wavefunction over all of spacetime. The equation is exceedingly complicated and cannot be solved completely.
Furthermore, in this theory time is not defined so there is no time evolution of the system. This is known as the
problem of time in quantum gravity. The canonical quantization approach also has not made any definite prediction
about the quantum nature of gravity in physical processes.\\

\vspace{.2in}
\noindent
{\bf 4 \hspace{.1in} The End Of Higher Dimensions}\\

\noindent
A remarkable paper recently revealed the severe challenges for gravity theories in higher dimensions and
effectively marked the end of Kaluza-Klein theory [7,8] after 90 years! We are therefore led to abandon an
elegant and long-held idea which nonetheless does not correspond to nature after thorough investigation. 
The work of Eingorn and Zhuk showed that all multidimensional gravity theories are found to be incompatible
with solar system observations [9]. For point-like masses, these theories cannot produce the gravitational
field which corresponds to known classical gravitational tests. Remarkably, the total number of spatial
dimensions $D$ is carried all the way in the equations of motion in four dimensional spacetime after 
compactification. When the equations of motion are applied to the classical tests in general relativity,
only the ordinary three dimensional case $(D = 3)$ agrees with observations. The result is independent of 
the size of the extra dimensions as long as they are compact and have the geometry of tori. For the classical
tests of these higher dimensional theories, a nonrelativistic weak field approximation is sufficient for the
treatment of motion in the solar system. The metric coefficients are obtained to lowest order in $1/c^{2}$ 
and they inevitably contain the total spatial dimensionality $D$ in their expressions; $c$ is the speed of light.\\

\noindent
The classical tests of general relativity considered are the following:\\

\noindent
1. Gravitational frequency shift.\\
2. Perihelion shift of planets.\\
3. Deflection of light.\\
4. Radar echo delay.\\
5. Parameterized post-Newtonian (PPN) parameters.\\

\noindent 
In the first case, the gravitational frequency shift, there is no observable difference between general relativity
and Kaluza-Klein theory in the weak field limit because the two results coincide in the lowest order. For all 
other tests: perihelion shift of the planet Mercury; bending of light by the Sun; radar echo delay from the 
Cassini spacecraft experiment; parameterized post-Newtonian parameters; the predictions from higher dimensional
theories $(D>3)$ in the lowest order approximation are significantly incompatible with known observed data. 
The calculation shows that larger spatial dimensionality always gives lower predicted values. All multidimensional
gravity theories involving equations of motion derived from the Hilbert action in higher dimensions, including
Kaluza-Klein theories, supergravity theories and superstring theories, face insurmountable challenge from solar
system data.\\

\vspace{.2in}
\noindent
{\bf 5 \hspace{.1in} Is Quantum Gravity An Illusion?}\\

\noindent
Despite intensive efforts to create a quantum theory of gravity, the goal is still illusive. We are no closer to
the goal than 30 years ago as more and serious difficulties have developed [10]. Many imaginative theories have been
proposed; none is yet successful and some have met remarkable defeat [11]. None of them has offered any prediction
in strong gravity environment or answered any question in physics. We are led to ponder if all such efforts are futile
because gravity is intrinsically classical and there is no need for a quantum theory of gravity. The situation may be
like the Navier-Stokes equation in fluid mechanics which is only a classical theory [12]. We believe that all
fundamental interactions are quantum in nature as shown by the success of those quantum field theories for particle
interactions, but a derived interaction may be classical. An excellent case is nuclear physics. It is well known that
nuclear force is not a fundamental force but an effective interaction from an underlying theory which is quantum
chromodynamics, constructed in terms of quarks and gluons. When a quantum field theory of nuclei was first attempted
the resulting theory would become inconsistent despite initial agreement with experiment [13]. A new structure at
a deeper level is needed.\\

The investigation of microscopic black holes can further shed light on the nature of gravity at the smallest scale.
Quantum black holes are intrinsically semi-classical objects rather than fully quantum objects [14]. They may require
treatment by quantum mechanics but not necessarily quantum field theory for their description. The difference is in
the pair creation and annihilation of virtual particles. A black hole of the size of an atomic nucleus has a mass of
a billion tons! Does quantum field theory really apply to a system with such a provocative mass? Are there creation
and annihilation operators for black holes in a Hilbert space? Where is the energy to create such a system? In 
quantum field theory, renormalization effects involving virtual particles still require the use of the {\it physical
mass} for these particles; in this case, a mass of a billion tons. When a particle has a mass much higher than the 
physical scale of interest, renormalization effects become negligible and the heavy particle decouples from the
theory. These problems do not exist if quantum black holes are only semi-classical objects.\\

The observation of the highest energy gamma rays up to 31 GeV from a distant gamma ray burst GRB 090510 also shows
that there are no observable quantum effects of spacetime down to the Planck scale [15]. The result therefore rules out
those quantum gravity theories in which the speed of light varies linearly with photon energy. There is no evidence
of violation of Lorentz invariance down to the Planck length. Spacetime is continuous and special relativity is right.
The greatest mystery is why spacetime is manifestly so smooth and classical all the way to the smallest conceivable
level.\\

\vspace{.2in}
\noindent
{\bf 6 \hspace{.1in} An Underlying Theory For Gravity}\\

\noindent
A new direction to understand gravity has recently been explored by considering classical gravity to be a {\it derived}
interaction from an underlying theory. This underlying theory would involve new degrees of freedom at a deeper level
and it would be structurally different from classical gravitation. It may conceivably be a quantum theory or a 
non-quantum theory, but more importantly, Einstein's gravity is not the $\hbar = 0$ limit of this theory. The relation
between this underlying theory and Einstein's gravity is like that of statistical mechanics and thermodynamics. 
In statistical mechanics, the fundamental object of a system is the partition function\\

\begin{equation}
  Z = \sum_{j} e^{-\beta E_{j}}  
\end{equation}
Here $\beta = 1/kT$, $k$ is the Boltzmann constant, $T$ is the temperature, and $E_{j}$ are the energy levels.  
The system is defined in a volume $V$. The thermodynamic quantities such as internal energy $U$, pressure $P$,
and entropy $S$ are then derived by the connections:

\begin{equation}
  U = - \frac{\partial \ln Z}{\partial \beta} \\
\end{equation}

\begin{equation}
  P = \frac{1}{\beta} \frac{\partial \ln Z}{\partial V} \\
\end{equation}

\begin{equation}
  S = k \ln Z + \frac{U}{T} .\\
\end{equation}

\noindent
Thermodynamic quantities are macroscopic variables chosen for performing experiments. The underlying degrees of
freedom of the system are the microscopic atoms and molecules. They obey the laws of quantum mechanics or classical
mechanics. However, it is not possible to quantize the variables in thermodynamics to arrive at the laws of
statistical mechanics. We believe that gravity is in a similar situation. It would be unreasonable to quantize
Einstein's gravity to arrive at the underlying quantum gravity theory because the metric and the connections are
macroscopic variables. They are intended for large scale description of spacetime in the first place.\\

A surprising parallel between classical gravity and thermodynamics is seen if we take Einstein's equation to be
a macroscopic equation with its basic variables and compare it to the First Law of Thermodynamics, given in
standard notations:\\

\begin{equation}
  R_{\mu \nu} - \frac{1}{2} g_{\mu \nu}R = \frac{8 \pi G}{c^{4}} T_{\mu \nu}\\ 
\end{equation}
and\\
\begin{equation}
  \delta Q - \delta W = dU  .\\
\end{equation}

\noindent
Both are energy equations if the constants $c$ and $G$ in Einstein's equation are rearranged to the left hand side.
Both can apply to different conditions such as matter-free region $T_{\mu \nu} = 0$ for gravity and constant energy
processes $dU = 0$ in thermodynamics. Is there a transformation which can connect the thermodynamic terms in the
First Law of Thermodynamics to the geometric terms of spacetime in Einstein's equation? Such a derivation has been
given using the Bekenstein-Hawking relation [16,17] for a black hole between the entropy $S$ and the area $A$ of a
causal horizon, together with the basic thermodynamic definition of heat $\delta Q = T dS$ for every point in
spacetime. On the gravity side, $\delta Q$ would be the energy flux across the horizon and $T$, the Unruh 
temperature [18] associated with an observer stationary at the horizon. The result is that Einstein's equation
is indeed viewed as an equation of state, or as the thermodynamics of spacetime [19]. The identification can be
more general on the horizon of a wide class of theories, thereby connecting the gravitational field equations and 
the thermodynamic equation [20]. It would be most remarkable if such identification can be carried out using the
partition function and in terms of new degrees of freedom. The real `quantum gravity theory' would then correspond
to the statistical mechanics of spacetime.\\

In 1917, a year after the completion of general relativity, Einstein expressed the following opinion to Felix Klein [21]:
\begin{quotation}
\noindent
{\em ``However we select from nature a complex of phenomena using the criterion of simplicity, in no case will its
theoretical treatment turn out to be forever appropriate. Newton's theory, for example, represents the gravitational
field in a seemingly complete way by means of the potential $\phi$. This description proves to be wanting; the
functions $g_{\mu \nu}$ take its place. But I do not doubt that the day will come when that description, too, 
will have to yield to another one, for reasons which at present we do not yet surmise. I believe that this process
of deepening the theory has no limit."}
\end{quotation}
\noindent
What is the new description? What is the new concept? What are the new variables? Who can surmise?

\end{document}